# THE AUTOIGNITION OF CYCLOHEXANE AND CYCLOPENTANE IN A SHOCK TUBE

B. SIRJEAN, F. BUDA, H. HAKKA, P.A. GLAUDE, R. FOURNET,

V. WARTH, F. BATTIN-LECLERC*

Département de Chimie-Physique des Réactions,

UMR n°7630 CNRS, INPL-ENSIC,

1 rue Grandville, BP 451, 54001 NANCY Cedex, France

*Colloquium : Reaction kinetics*

**Total word count (Method 1) : 5571**

- Text : 3183
- References : (33+2)x2.3x7.6 = 612
- Table 1 : (5+2)x7.6x2 + 40 = 146
- Equations : (2+4)x7.6 = 46
- Figures : 1584
  - Figure 1 = (50+10)x2.2 + 15 = 147
  - Figure 2 = (135+10)x2.2 + 38 = 379
  - Figure 3 = (65+10)x2.2 + 83 = 248
  - Figure 4 = (40+10)x2.2 +41 = 151
  - Figure 5 = (45+10)x2.2 + 33 = 154
  - Figure 6 = (95+10)x2.2 + 33 = 212
  - Figure 7 = (110+10)x2.2 + 29 = 293

* E-mail : Frederique.battin-Leclerc@ensic.inpl-nancy.fr ; Tel.: 33 3 83 17 51 25 , Fax : 33 3 83 37 81 20

Ignition delay times of cyclohexane-oxygen-argon and cyclopentane-oxygen-argon mixtures have been measured in a shock tube, the onset of ignition being detected by OH radical emission. Mixtures contained 0.5 or 1 % of hydrocarbon for equivalence ratios ranging from 0.5 to 2. Reflected shock waves allowed temperatures from 1230 to 1800 K and pressures from 7.3 to 9.5 atm to be obtained. These measurements have shown that cyclopentane is much less reactive than cyclohexane, as for a given temperature the observed autoignition delay times were about ten times higher for the $C_5$ compound compared to the $C_6$.

Detailed mechanisms for the combustion of cyclohexane and cyclopentane have been proposed to reproduce these results. The elementary steps included in the kinetic models of the oxidation of cyclanes are close to those proposed to describe the oxidation of acyclic alkanes and alkenes. Consequently, it has been possible to obtain these models by using an improved version of software EXGAS, a computer package developed to perform the automatic generation of detailed kinetic models for the gas-phase oxidation and combustion of linear and branched alkanes and alkenes. Nevertheless, the modelling of the oxidation of cyclanes requires to consider new types of generic reactions, and especially to define new correlations for the estimation of the rate constants. Ab initio calculations have been used to better know some of the rate constants used in the case of cyclopentane. The main reaction pathways have been derived from flow rate and sensitivity analyses.

## INTRODUCTION

Cycloalkanes have a significant presence in conventional fuels (up to 3 % in gasoline and up to 35% in diesel fuel), but relatively little attention has been paid to the chemistry involved during their oxidation [1]. Furthermore, cycloalkanes are key compounds of the formation of aromatic pollutants.

Monocyclic small ring hydrocarbons do not appear to have been much studied. Ignition delay times have been reported for cyclopropane and cyclobutane in stoichiometric air-like argon mixtures between 1200 and 1600 K showing that cyclopropane is less reactive than cyclobutane [2].

An experimental study of the oxidation of cyclopentane has been performed at 873 K in a jet-stirred reactor at 53 kPa, for a rich mixture, with residence times from 0.1 to 0.5 s corresponding to fuel consumption between 2 and 24 % [3]. Autoignition delay times of methylcyclopentane have been measured in a shock tube for temperatures ranging from 1200 to 2200 K, pressures from 0.5 to 2 bar and equivalence ratios from 0.5 to 2 [4].

More papers are related to cyclohexane. Klai and Baronnet [5,6] have studied its oxidation in a static reactor at 635 K, 4.7 kPa, for an equivalence ratio equal to 9. More recently, El Bakali et al. [7] have studied the oxidation of cyclohexane in a jet-stirred reactor for temperatures ranging from 900 to 1150 K, pressures from 1 to 10 atm and equivalence ratios from 0.5 to 1.5. Laminar flame speeds of cyclohexane/air flames have been measured by Davis and Law [8]. Lemaire et al. [9] have determined autoignition delay times in a rapid compression machine for temperatures ranging from 650 to 900 K, pressures from 7 to 17 bar, for an equivalence ratio equal to 1.

As no previous work is related to the autoignition of cyclopentane or cyclohexane above 1000 K, we present here an experimental and modelling study of the autoignition of these two compounds in a shock tube.

## EXPERIMENTAL PROCEDURE AND RESULTS

The measurement of ignition delay times in the used shock tube has been described in several papers [10-11]; the main features of this experimental device will just be recalled here. The stainless steel shock tube included a reaction and a driver parts separated by two terphane diaphragms, which were ruptured by decreasing suddenly the pressure in the space separating them. The driver gas was helium. The incident and reflected shock velocities were measured by four piezo-electric pressure transducers located along the reaction section. The pressure and temperature of the test gas behind the reflected shock wave were derived from the values of the initial pressure in the low pressure section and of the incident shock velocity by using ideal one-dimensional shock equations. The onset of ignition were detected by excited OH* radical emission at 306 nm through a quartz window with a photomultiplier fitted with a monochromator at the end of the reaction part. The quartz window was located at the same place along the axis of the tube as the last pressure transducer. Figure 1 presents the record of the signal of the last pressure transducer and of the OH* emission for a typical experiment. The pressure profile displayed three rises, which were due to the incident shock wave, pointed as 1, the reflected shock wave pointed as 2, and the ignition. This last pressure rise is too weak to be used for determining the delay time, while the rise of OH emission, pointed as 3, is much steeper. The ignition delay time was defined as the time interval between the pressure rise measured by the last pressure transducer due to the arrival of the reflected shock wave and the rise of the optical signal by the photomultiplier up to 10% of its maximum value.

Cyclohexane was purchased from SDS, with a purity of 99.5% and cyclopentane from Sigma-Aldrich with a purity of 99%. Oxygen, argon and helium were purchased from Messer (purity > 99.995). Fresh reaction mixtures were prepared every day and mixed using a recirculation pump. Before each introduction of the reaction mixture, the reaction section was flushed with pure argon and evacuated, to ensure the residual gas to be mainly argon.

This study was performed in the following experimental conditions after the reflected shock :

Temperature range from 1230 to 1840 K,

Pressure range from 7.3 to 9.5 bar,

Mixtures (argon /cyclohexane / oxygen, in molar percent) were (90.5 / 0.5 / 9), (95 / 0.5 / 4.5), (97.25 / 0.5 / 2.25) and (90 /1 / 9) and mixtures (argon /cyclopentane / oxygen, in molar percent) were (92 /0.5 / 7.5), (95.75 / 0.5 / 3.75), (97.625 / 0.5 / 1.875) and (91.5 / 1 / 7.5), respectively, corresponding to three different equivalence ratios ($\Phi$ = 0.5, 1 and 2) and to two different concentrations of hydrocarbon (0.5 and 1 %) and leading to delay times from 6 up to 2190 μs. Figures 2 presents the experimental results obtained for the ignition of cyclohexane and cyclopentane, in the case of the mixture containing 0.5 % of hydrocarbon, for three equivalent ratios. These results show that in each case, ignition delay times increase when equivalence ratio increases.

An additional (argon /cyclopentane / oxygen) mixture of (95 / 0.5 / 4.5), with $\Phi$ = 0.83, has also been studied to allow the direct comparison with the same mixture with cyclohexane presented in Figure 3 and showing a strong difference of reactivity between $C_5$ and $C_6$ compounds, as the ignition delay times for cyclopentane are almost 10 times longer than for cyclohexane for the lowest temperatures studied. This difference is less marked when temperature increases.

## DESCRIPTION OF THE DETAILED KINETIC MODEL

The detailed chemical kinetic reaction mechanisms presented here have been developed with the help of an improved version of software EXGAS and can be obtained on request.

As the use of EXGAS to model the oxidation of alkanes and alkenes has already been extensively described [12-15], we will recall here only its main features and we will further detail the specificities concerning cyclanes.

*General features of EXGAS*

The system provides reaction mechanisms made of three parts :

- A comprehensive primary mechanism, in which the only molecular reactants considered are the initial organic compounds and oxygen. The primary mechanism to model the autoignition of cyclanes at high temperatures includes the following elementary steps:
  - Unimolecular and bimolecular initiation steps,
  - Decomposition by beta-scission and oxidation (to form the conjugated alkene plus $HO_2$•) of alkyl radicals,
  - Isomerizations of alkyl radicals,
  - Metatheses involving the H-abstraction reactions from the initial reactants by small radicals.
- A $C_0$-$C_2$ reaction base, including all the reactions involving radicals or molecules containing less than three carbon atoms [16], which is periodically updated, which is coupled with a reaction base for $C_3$-$C_4$ multi-unsaturated hydrocarbons [10, 11], such as propyne, allene or butadiene, including reactions leading to the formation of benzene and in which pressure-dependent rate constants are considered. Most $C_0$-$C_2$ and many $C_3$-$C_4$ pressure-dependent rate constants follow the Troe's formalism and efficiency coefficients have been included.

- A lumped secondary mechanism, containing reactions consuming the molecular products of the primary mechanism which do not react in the reaction bases [12]. No secondary mechanism is written for dienes.

Thermochemical data for molecules or radicals are automatically computed using software THERGAS [17], based on group additivity [18], and stored as 14 polynomial coefficients, according to the CHEMKIN II formalism [19].

The kinetic data of isomerizations, recombinations and the unimolecular decompositions are calculated using software KINGAS [12] based on the thermochemical kinetics methods [18]. The kinetic data, for which the calculation is not possible by KINGAS, are estimated from correlations, which are based on quantitative structure-reactivity relationships and obtained from a literature review. The main features of these calculations and estimations have been summarized in previous descriptions of EXGAS [12-13].

*Specificities concerning the modeling of the oxidation of cyclanes*

On a computing point of view, the treatment of cyclic reactants has induced important modifications, as a specific internal notation had to be taken into account with a new algorithm of canonicity to avoid redundant products or generated reactions [20].

On a chemical point of view, the reaction of cyclanes at high temperature are close to that of alkanes. The main differences lies in unimolecular initiation steps involving the formation of a biradical by breaking of a C-C bond. This initiation and the subsequent reactions have been proposed by Tsang to explain results obtained during the pyrolysis of cyclopentane [22] and cyclohexane [23] in a single pulse shock tube and by Orme et al to model the autoignition of methylcyclopentane [4].

This initiation reaction and the subsequent reactions, which we have considered here in the case of cyclohexane, are presented in Figure 4. Tsang [22] has proposed that $k_1/k_{-1}$= 4.0x10$^6$ exp(-41800/T) and $k_2$ = 1.26x10$^{10}$ exp(-2600/T) s$^{-1}$ and to consider the same ratio between the rates of the disproportionnation and of the combination than for propyl radicals; i.e. 0.16. These assumptions has led us to use $k_1 = 3.1.x10^{17} exp(-2600/T)\ s^{-1}$ and $k_{-1} = 7.9x10^{10} exp(-45000/T)\ s^{-1}$. The β-scission decompositions (4) and (5) involve the breaking of a C-C bond and have then a rate constant $k_4$ = $2k_5 = 4x10^{13}exp(-14490/T)\ s^{-1}$, i.e. the double of the value for an alkyl radical for reaction (4) in which 2 C-C bonds can be broken. Reaction (3) is a beta-scission decomposition involving the breaking of a C-H bond. By analogy with alkyl radicals, its rate constant is $k_3 = 6x10^{13}exp(-19190/T)\ s^{-1}$, since 4 C-H bonds can be broken.

For cyclopentane, the direct formation of 1-pentene (reaction 1') and of cyclopropane and ethylene (reaction 2') has been written without considering the intermediate formation of a biradical. The rate constant are $k_{1'} = 1.25x10^{16}\ exp(-42850/T)\ s^{-1}$ and $k_{2'} = 1.77x10^{16}exp(-48000/T)\ s^{-1}$, as proposed by Tsang [22]. For both compounds, the rate constant, $k_6$, of the unimolecular decomposition to give H atoms and cycloalkyl radicals have been deduced from that of the reverse reaction, $k_{-6} = 1.0\ x10^{14}$ $s^{-1}$ according to Allara et al. [23].

The rate constants for the bimolecular initiation, the decomposition by β-scission, the oxidations and the metatheses of cyclic compounds have been taken equal to that of the similar reaction for alkanes and alkyl radicals [13]. Only the A-factors for the metatheses by H-atoms from cycloalkanes to give cycloalkyls radicals have been considered five times higher than in the case of alkanes ($k' = 2.70x10^8T^2exp(-5000/T)\ cm^3.mol^{-1}.s^{-1}$ and $k_{7'} = 2.25x10^8T^2exp(-5000/T)\ cm^3.mol^{-1}.s^{-1}$); with this change the rate constants used here are in agreement with the values measured by the team of Walker [24, 25] within a factor around 2. In addition, the decompositions by β-scission involving the breaking of a C-H bond from cyclohexyl radicals to give cyclohexene (8) is taken into account as the reverse of the corresponding additions of H atoms to the double bond. As in our previous work concerning the oxidation of alkenes, the rate constant for this addition is $k_{-8} = 2.6x10^{13}exp(-790/T)\ s^{-1}$ [26].

The decompositions by a β-scission involving the breaking of a C-C bond from cyclopentyl radicals (9') induces the opening of a ring with a strain energy, while it is not the case for cyclohexane. To better consider this reaction and the concurrent step, the beta-scission involving the breaking of a C-H bond (8') and leading to cyclopentene, we have calculated the related rate constants using ab initio calculations. These calculations have been performed with the CBS-QB3 method of Gaussian03 [27]. The choice of this composite method is justified by the accuracy sought in kinetic calculations and by the low number of heavy atoms involved in our system. Explicit treatment of internals rotors has been performed with the *hindered Rotor* option of Gaussian03 in accordance with the work of Ayala et al. [28]. Frequency analyses have made it possible to point out one imaginary frequency for each transition state (TS). Moreover, Intrinsic Reaction Coordinate (IRC) calculations have been systematically performed at the B3LYP/6-31G(d) level on TS, to ensure that they are correctly connected to the desired reactants and products. The high-pressure limit unimolecular rate constants involved in the β-scission were calculated using transition state theory [29] :

$$k_{uni} = \frac{k_b T}{h} \exp\left(\frac{\Delta S^{\neq}}{R}\right) \exp\left(-\frac{\Delta H^{\neq}}{RT}\right) \qquad \{1\}$$

where $\Delta S^{\neq}$ and $\Delta H^{\neq}$ are, respectively, the entropy and enthalpy of activation. The kinetic parameters were obtained from a fitting of equation {1}, for several temperatures, with a modified Arrhenius forms (equation {2}):

$$k = A\, T^{b} \exp(-E/RT) \qquad \{2\}$$

Figure 5 presents the corresponding activation enthalpies and free enthalpies calculated at the CBS-QB3 level at 1300K**.** The rate constants used for the β-scission of cyclic radicals are summarized in Table1. The kinetic parameters, for the β-scission of the C-C bond of cyclopentyl radical, can be compared with the rate constants obtained in a low temperature range by Handford-Styring and Walker [25] by fitting a complex mechanism from experimental results ($k = 1.4\times10^{13}\exp(-17260/T)$ $s^{-1}$) and by Gierczak et al. from RRKM extrapolation [30] ($k = 2.4\times10^{14}\exp(-16100/T)$ $s^{-1}$). At 600K, the rate constant obtained by Giecrzak et al is about 100 times higher than that obtained by Handford-Styring and Walker. The activation energy obtained from our calculations (reaction 9') is close to that proposed by Handford-Styring and Walker, while our rate constant at 600K is ten times higher and then intermediate between this one and that proposed by Gierczak. Another point concerns the relatively high activation energy for this β-scission (reaction 9') (in comparison, Ea/R = 14443 for an alkyl radical [13]), if we keep in mind the ring strain of cyclopentane of about 6.3 kcal/mol [18]. As previously mentionned by Stein and Rabinovitch [31], two opposite effects may alter ring opening energetics from acyclic values. The first one is the ring strain energy recovery if the transition state (TS) is loosened, which tends to decrease activation energy ($E_a$). The second one is due to some local orientation strain which tends to increase $E_a$. The TS corresponding to the reaction 9' is shown in figure 6a ; the length between the two carbon atoms involved in the breaking bond is only 1.45 times the corresponding distance in cyclopentyl radical (figure 6b). ~~This value can be compared with the factor 2.5 estimated by Benson for unimolecular initiations [18].~~ So, the ~~calculated~~ TS is relatively tight and has a structure close to that of the initial cyclic reactant. Then a part of the ring strain energy remains in the TS and cannot be removed from the activation energy. Moreover, the formation of a nascent π bond in the TS involved a steric inhibition as shown in figure 6a. These two effects could explained the high activation energy obtained.

A schematic mechanism for the oxidation of cyclopentene has been written and includes initiations, metatheses with H-atoms and OH radicals and the decomposition by β-scission by breaking of a C-H bond,. The mechanisms recently published for the oxidation of cyclopentadiene [31] and cyclohexene [32] have been added in order to well consider the secondary reactions of these molecules. To take into account the reactions of the 1-alkenes obtained by the unimolecular initiations, the mechanisms recently published for the oxidation of 1-pentene [14] and 1-hexene [15] have been added.

**DISCUSSION**

Simulations were performed using software SENKIN of CHEMKIN II [19]. Figures 2 and 3 display a comparison between our experimental results and ignitions delay times computed using the mechanisms previously described, which globally contain 1629 reactions in the case of cyclohexane (789 reactions in the reactions bases, 340 in the primary and 500 in the secondary mechanisms) and 1204 reactions in the case of cyclopentane (211 in the primary and 284 in the secondary mechanisms). Simulated ignitions delay times have been taken as the time at 10 % of the maximum concentration of OH radicals. The same results are obtained when considering the rise of CH radicals. The reaction of CH radicals with oxygen is usually considered as the main source of excited OH* radicals responsible for the emission at 306 nm [33]. The agreement between experiments and simulation is globally acceptable and, as shown in Figure 3, the mechanisms reproduces well the difference of reactivity between the $C_6$ and $C_5$ compounds.

Figure 7 displays flow rate analyses performed for cyclohexane and cyclopentane. While a small amount of 1-alkenes is obtained by initiation steps, both cyclanes react mainly by H-abstraction with hydrogen atoms and hydroxyl radical to form the corresponding cycloalkyl radicals. The fate of these cyclic radicals are different according to the size of the ring. Cyclopentyl radicals give mainly 1-penten-5-yl radicals, while cyclohexyl radicals lead in almost equal proportions to 1-hexen-6-yl radicals and to cyclohexene and hydrogen radicals. Hydrogen radicals are involved in the branching step with oxygen molecules to give oxygen atoms and hydroxyl radicals. Their formation promotes the multiplication of radicals and explains then why cyclohexane is more reactive than cyclopentane. As shown in Table 1, the difference in reactivity does not come from differences in the rate constants of the direct decomposition by β-scissions, but it is due to thermochemistry. The difference in reaction free enthalpies shown in Table 1 makes the β-scission involving the breaking of a C-H bond more easily equilibrated in the case of cyclopentane than for cyclohexane.

According to figure 7a, 1-hexen-6-yl radicals are consumed by a beta-scission decomposition to give ethylene and 1-buten-4-yl radicals and by isomerisation to give 1-hexen-3-yl radicals. 1-buten-4-yl radicals leads to ethylene and vinyl radicals or after isomerization to acetylene and ethyl radicals. Resonance stabilized 1-hexen-3-yl radical decompose to give 1,3-butadiene and ethyl radicals. According to figure 7b, 1-penten-5-yl radicals are rapidly consumed to give ethylene and allyl radicals which lead to propene and to acroleine.

Figure 3 presents a sensitivity analysis on the initiation steps proposed by Tsang [21, 22] and shows that these reactions have a slight promoting effect in the case of both cyclohexane and cyclopentane. This promoting effect observed for both cyclanes is due to the easier unimolecular initiations from 1-alkenes to give ethyl (in the case of 1-pentene) or propyl (in the case of 1-hexene) and resonance stabilized allyl radicals ($E_a$ = 71 kcal/mol) compared to initiations from cyclanes by breaking of a C-H bond ($E_a$ = 99.5 kcal/mol). The promoting effect is more important in the case of cyclopentane, because ethyl radicals can easily lead to the formation of hydrogen atoms, which are involved in the branching step, $H+O_2 \rightarrow OH+O$, which is usually the most sensitive parameter in our modeling of ignition delay times in a shock tube [10-11].

Figure 3 shows also the important inhibiting effect of the metatheses with H-atoms, which consume H-atoms to produce less reactive radicals and are then in concurrence with the branching step, $H+O_2 \rightarrow OH+O$. This inhibiting effect is more important in the case of cyclohexane, because more H-atoms are obtained from the decomposition by β-scission from cyclochexyl radicals than from cyclopentyl radicals.

## CONCLUSION

This paper presents new measurements concerning ignition delay times of cyclohexane and cyclopentane in a shock tube between 1230 to 1800 K, as well as detailed kinetic models allowing these results to be correctly reproduced in most studied conditions.

The experimental results shows that the reactivity of cyclopentane is much lower than that of cyclohexane. Flow rate analyses indicate that this difference is mainly due to the difference of the fate of the corresponding cycloalkyl radicals: cyclopentyl radicals give mainly 1-penten-5-yl radicals, while cyclohexyl radicals lead in almost equal proportions to 1-hexen-6yl radicals and to cyclohexene and hydrogen radicals, the formation of which promotes the branching step, $H+O_2 \rightarrow OH+O$, and the multiplication of radicals.

## ACKNOWLEDGEMENTS

Financial support of this work by the European Union within the "SAFEKINEX" Project EVG1-CT-2002-00072 is gratefully acknowledged.

**FIGURE CAPTIONS**

Figure 1: Typical experimental profiles of pressure and OH emission in the case of cyclopentane.

Figure 2: Ignition delay times of (a) cyclohexane and (b) cyclopentane in shock tube for equivalence ratios of 0.5, 1 and 2 and a concentration of hydrocarbon of 0.5%. Points correspond to experimental results and lines to simulations.

Figure 3: Comparison between autoignition delay times of cyclopentane and cyclohexane at 0.5% hydrocarbon and of 4.5% oxygen (corresponding to an equivalence ratio of 1 for cyclohexane). Points correspond to experimental results, full lines to simulations with the complete mechanism, small broken lines to simulations with mechanisms in which the initiation steps proposed by Tsang [21, 22] have been removed and large broken lines to simulations with mechanisms in which the rate constant of metathesis with H-atoms has been divided by 5.

Figure 4: Initiation step and subsequent reactions involving the formation of biradicals.

Figure 5: Relative enthalpies and free enthalpies (in bold) of stable species and transition states calculated at the CBS-QB3 level in $kcal.mol^{-1}$ at 1300 K for the decompositions by beta-scission of cyclopentyl radicals.

Figure 6: Structures of the transition state involved in the β-scission of the C-C bond (a) and of cyclopentyl radical (b) at the CBS-QB3 level. The length of the bonds are given in Angstrom.

Figure 7: Flow rate analyses at $\Phi$=1 at 1360 K and 7.5 bar for 50% consumption of reactant for a mixture containing (a) 0.5% cyclohexane and (b) 0.5% cyclopentane.

Table 1 : Rate constants and reaction free enthalpies (at 1300 K) for the decomposition by β-scission of cyclic radicals in kcal, mol, $cm^3$, s units ($k_d$: rate constant of the direct reaction, $k_r$: rate constant of the reverse reaction).

| | Reactions | | A | n | Ea | $k_d$ | $\Delta G_r$ | $k_d/k_r$ |
|---|---|---|---|---|---|---|---|---|
| $C_5$ | c-$C_5H_9$=H+c-$C_5H_8$ | (8') | $4.4\times10^{14}$ | 0.104 | 37.3 | $4.8\times10^{8}$ | 10.30 | $1.4\times10^{-7}$ |
| | c- $C_5H_9$=l-$C_5H_9$ | (9') | $1.6\times10^{14}$ | 0.034 | 34.1 | $3.8\times10^{8}$ | 2.05 | 0.45 |
| $C_6$ | c-$C_6H_{11}$=H+c-$C_6H_{10}$ | (8) | $2.5\times10^{14}$ | 0 | 34.0 | $4.9\times10^{8}$ | -4.86 | $3.4\times10^{-5}$ |
| | c-$C_6H_{11}$=l-$C_6H_{11}$ | (9) | $2.6\times10^{13}$ | 0 | 28.7 | $6.0\times10^{8}$ | -3.98 | 4.4 |

Figure 1

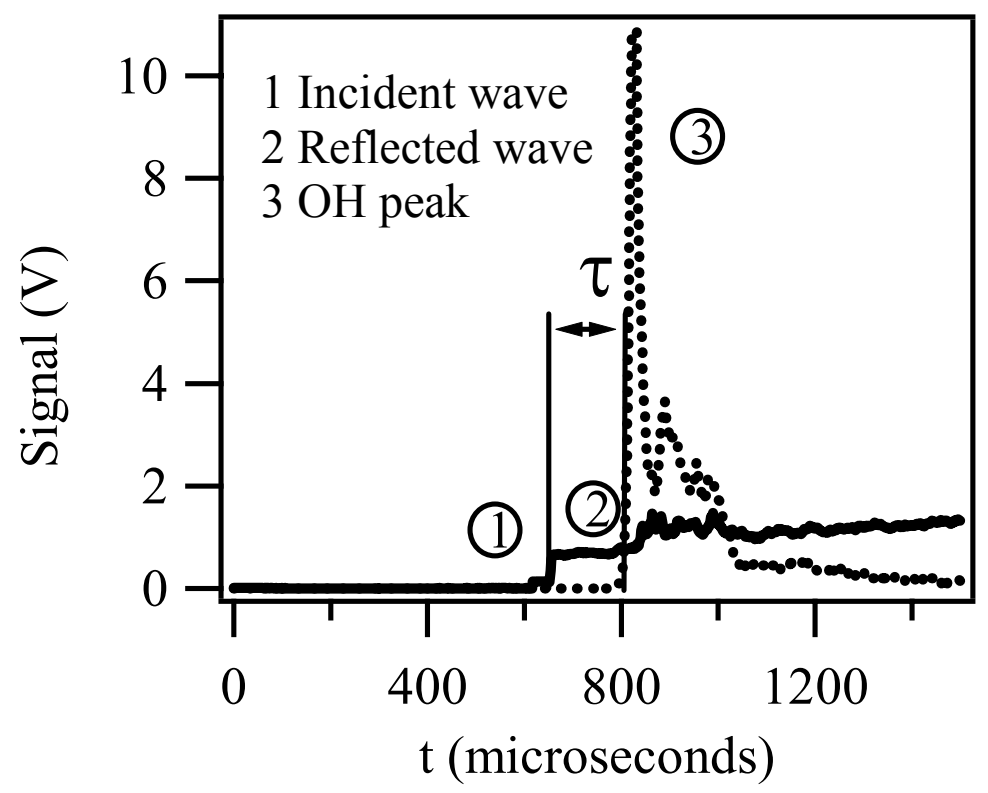

Figure 2

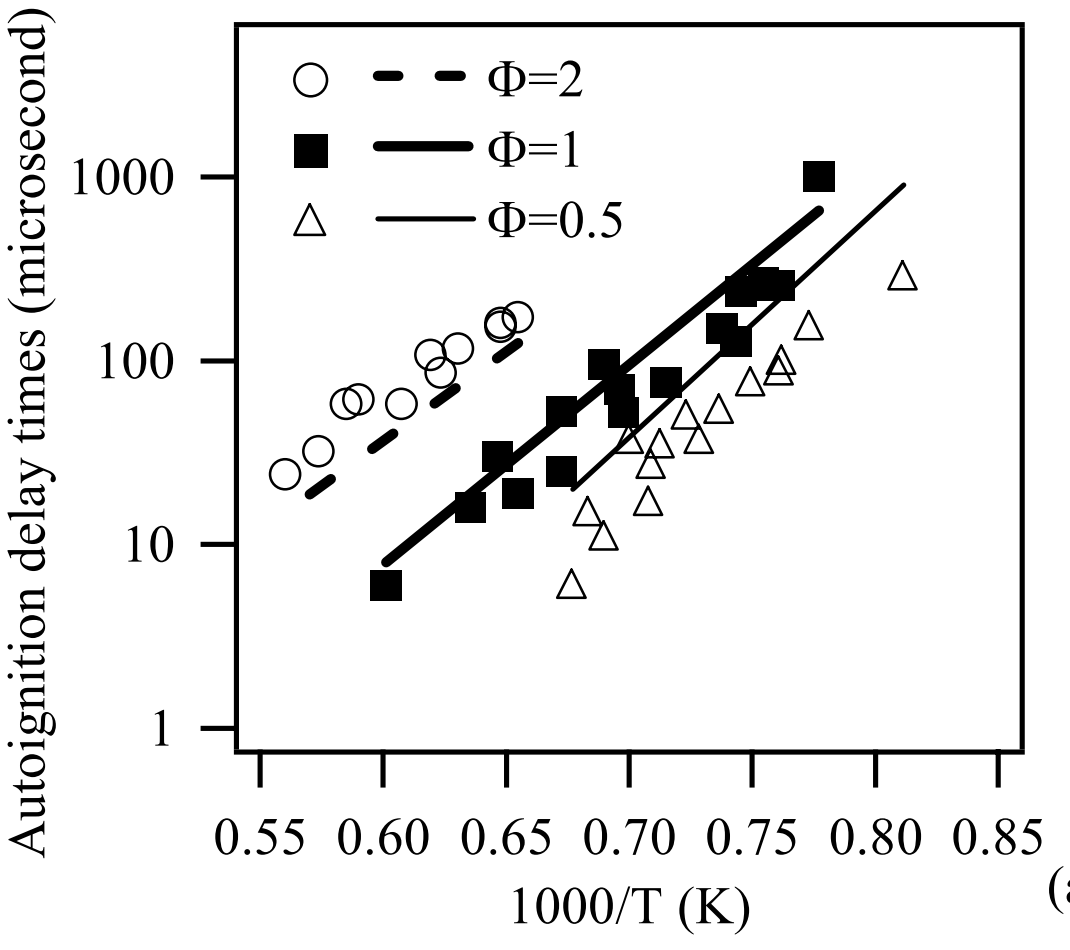

(a)

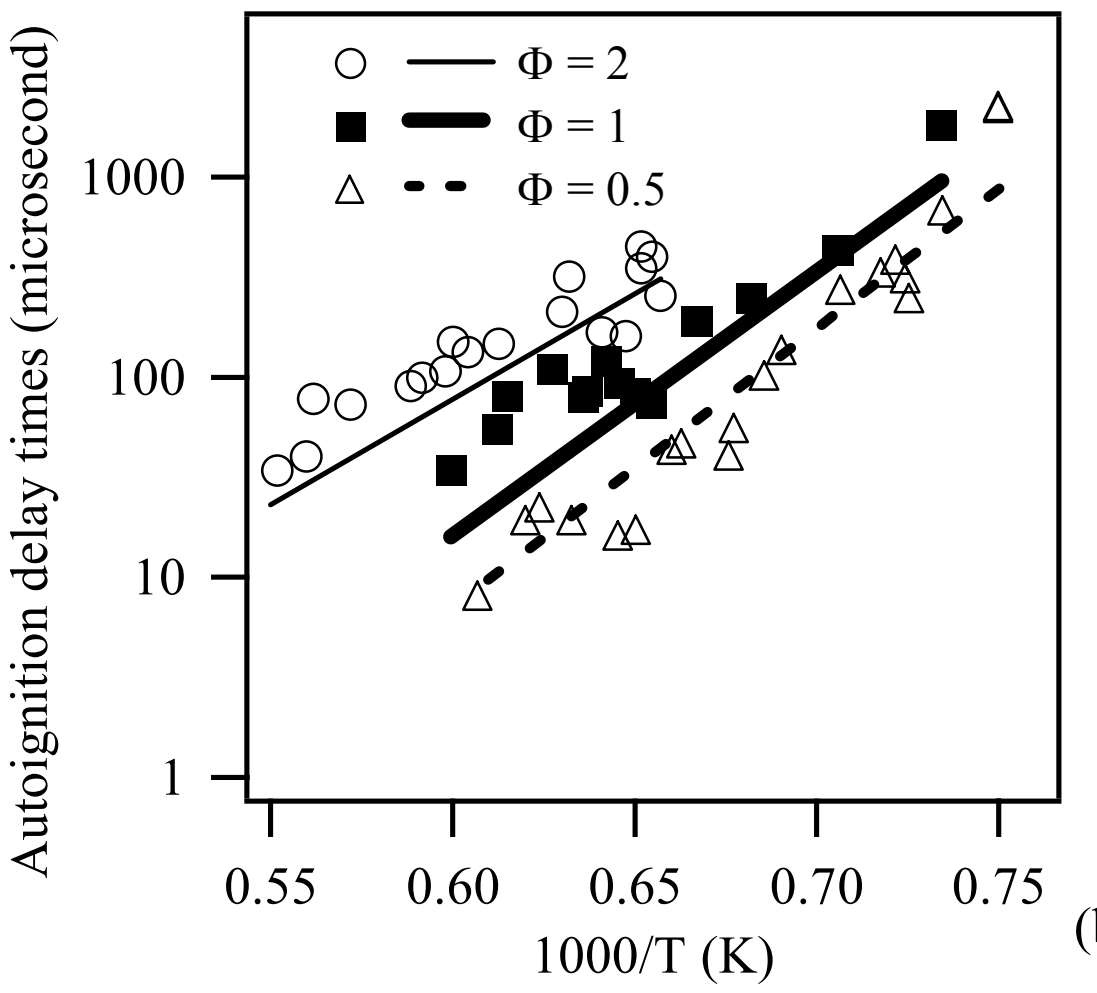

(b)

Figure 3

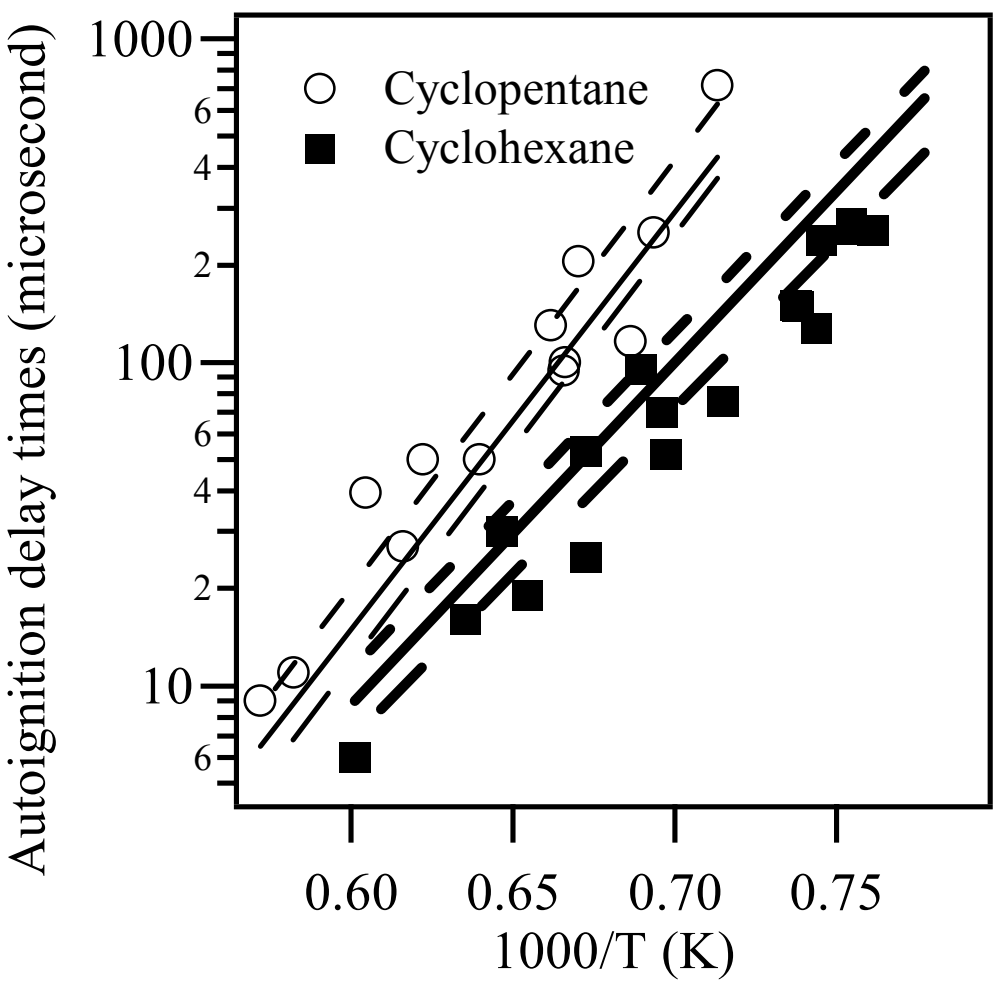

Figure 4

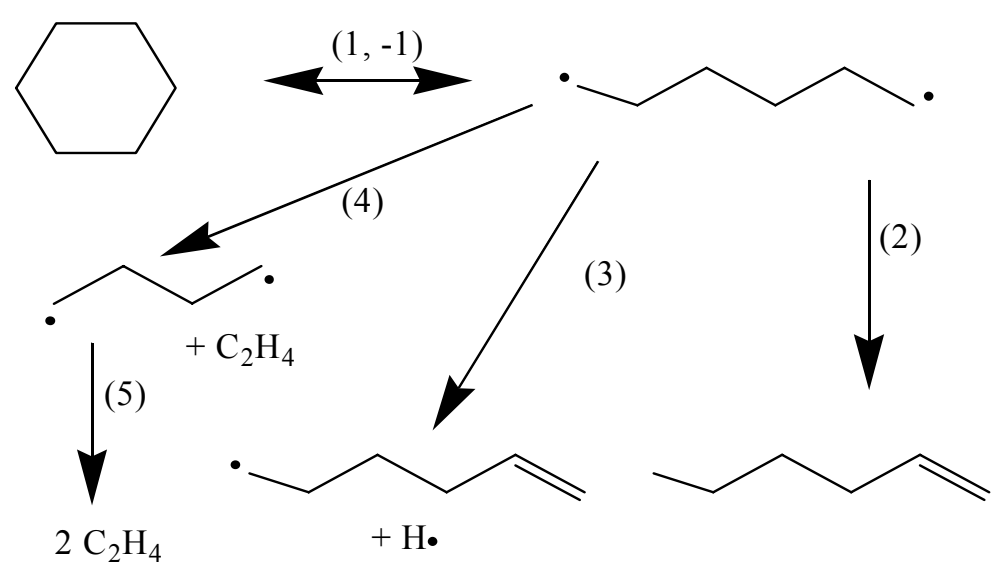

Figure 5

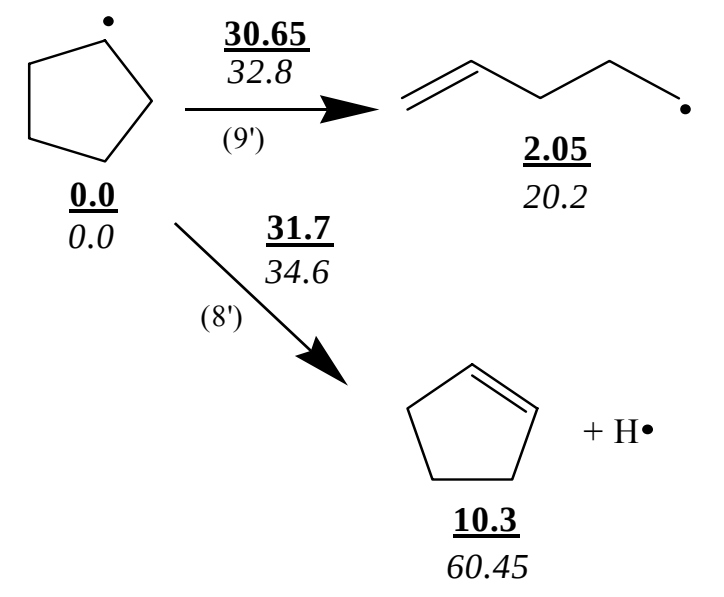

Figure 6

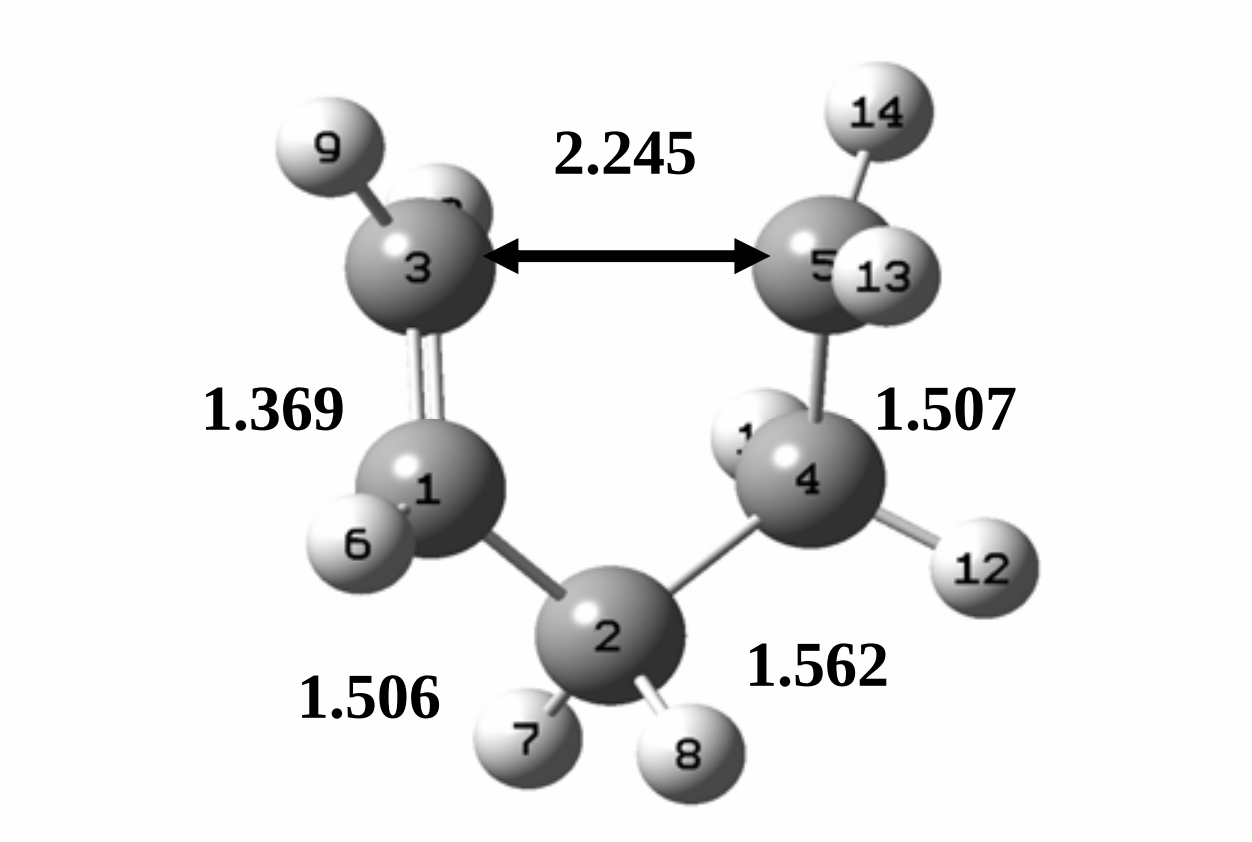

(a)

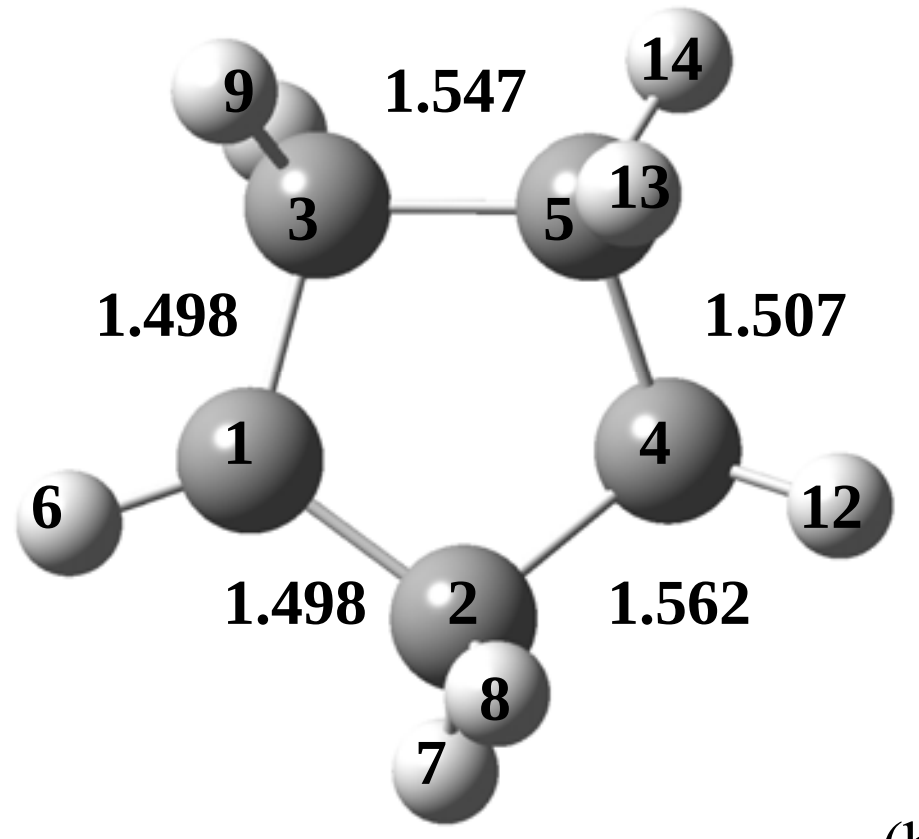

(b)

Figure 7

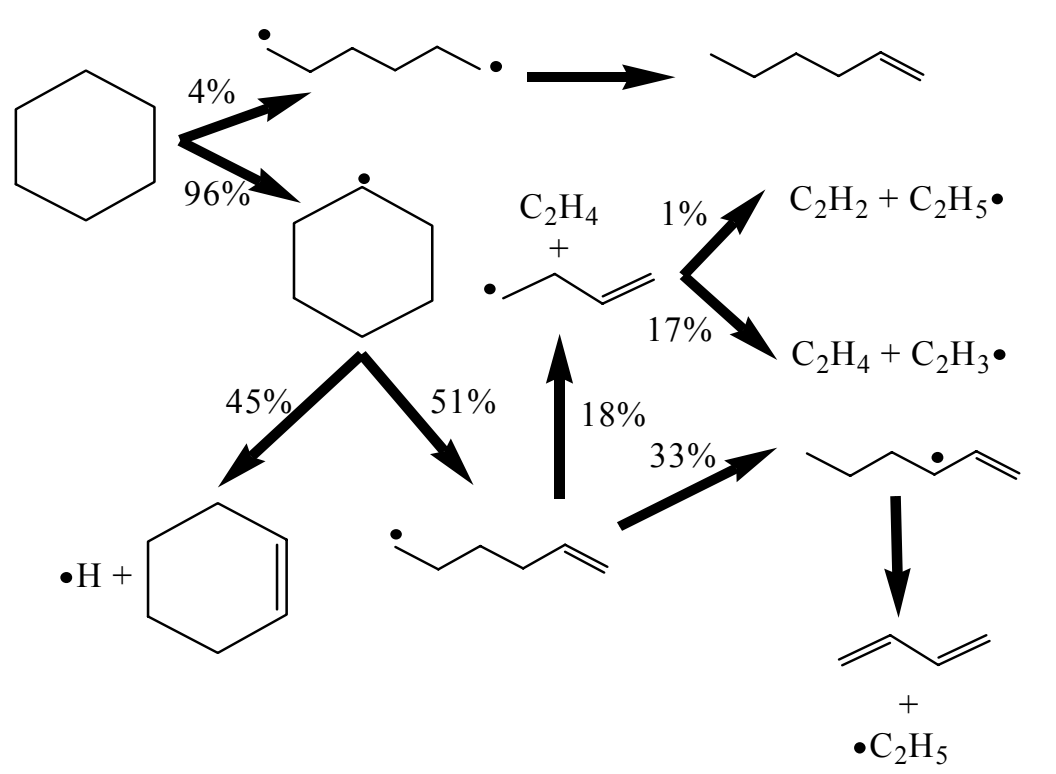

(a)

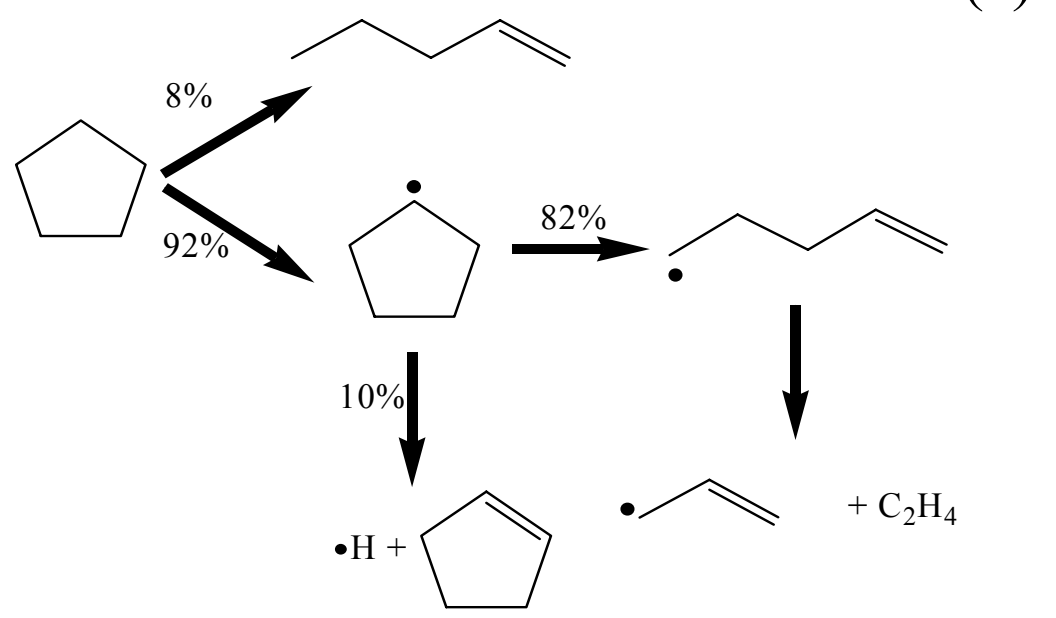

(b)